\documentclass[preprint,superscriptaddress,nofootinbib,dvipdfmx]{revtex4-1}
\usepackage{graphicx}
\usepackage{amsmath,amssymb,color,colortbl}

\newcommand{\red}[1]{{\color[rgb]{1,0,0} #1}}

\newcommand{\GeV}{{\text{GeV}}}
\newcommand{\TeV}{{\text{TeV}}}

\newcommand{\fb}{{\text{fb}}}

\newcommand{\BR}{\text{BR}}

\newcommand{\U}{{\text{U}}}
\newcommand{\SU}{{\text{SU}}}

\newcommand{\ellp}{\ell^\prime}
\newcommand{\dcheckmark}{\checkmark\!\!\checkmark}

\begin{document}

\preprint{KANAZAWA-16-10}
\preprint{UT-HET 115}

\title{
Testing neutrino mass generation mechanisms\\
from the lepton flavor violating decay of the Higgs boson
}

\author{Mayumi Aoki}
\email{mayumi@hep.s.kanazawa-u.ac.jp}
\affiliation{
Institute for Theoretical Physics,
Kanazawa University,
Kanazawa 920-1192, Japan
}
\author{Shinya Kanemura}
\email{kanemu@sci.u-toyama.ac.jp}
\affiliation{
Department of Physics,
University of Toyama,
3190 Gofuku,
Toyama 930-8555, Japan
}
\author{Kodai Sakurai}
\email{sakurai@jodo.sci.u-toyama.ac.jp}
\affiliation{
Department of Physics,
University of Toyama,
3190 Gofuku,
Toyama 930-8555, Japan
}
\author{Hiroaki Sugiyama}
\email{sugiyama@sci.u-toyama.ac.jp}
\affiliation{
Department of Physics,
University of Toyama,
3190 Gofuku,
Toyama 930-8555, Japan
}


\begin{abstract}

 We investigate how observations
of the lepton flavor violating decay of the Higgs boson%
~($h \to \ell\ellp$)
can narrow down models of
neutrino mass generation mechanisms,
which were systematically studied
in Refs.~\cite{Kanemura:2015cca, Kanemura:2016ixx}
by focusing on the combination of
new Yukawa coupling matrices with leptons.
 We find that
a wide class of models for neutrino masses can be excluded
if evidence for $h \to \ell\ellp$ is really obtained
in the current or future collider experiments.
 In particular,
simple models of Majorana neutrino masses
cannot be compatible with the observation of $h \to \ell\ellp$.
 It is also found that some of the simple models
to generate masses of Dirac neutrinos radiatively
can be compatible with a significant rate of the $h \to \ell\ellp$ process.

\end{abstract}

\maketitle

\section{Introduction}

 Since the discovery of the neutrino oscillation~\cite{ref:nu-osc},
the origin of small masses for neutrinos has been
one of the most important problems of particle physics.
 It would be rather unnatural
if the origin of such tiny neutrino masses
is the same as the one for quark and charged lepton masses.
 Therefore,
it would be expected that
neutrinos obtain masses via a different mechanism
from quarks and charged leptons.

 There can be two types of the mass for neutrinos;
e.g., Majorana masses and Dirac masses,
where the former break the lepton number conservation
by two units.
 There are simple scenarios to produce
Majorana neutrino masses at the tree level
by the seesaw mechanism.
 In the type-I~\cite{ref:seesaw, Schechter:1980gr},
II~\cite{ref:HTM, Schechter:1980gr},
and III~\cite{Foot:1988aq} seesaw scenarios,
the origin of the lepton number violation~(LNV)
is the mass of heavy right-handed neutrinos,
the scalar coupling with an $\SU(2)_L$-triplet Higgs field,
and the mass of triplet fermions, respectively.
 As an alternative scenario,
neutrino masses are generated at the loop level.
 The smallness of neutrino masses
can be explained not only by the large mass scale
but also by the loop suppression factor
and new coupling constants
which would be less than unity.
 The first model along this line
has been proposed by A.~Zee~\cite{Zee:1980ai},
in which neutrino masses are generated at the one-loop level
by introducing an extended Higgs sector.
 Subsequently,
many variant models have been proposed so far.
 For example,
there are models
where neutrino masses are generated
at the one-loop or higher-loop levels%
~\cite{Zee:1985id, Babu:1988ki, Cheng:1980qt,
ref:Ma, ref:KNT, ref:AKS, ref:GNR},
some of which involve the dark matter candidate running in the loop%
~\cite{ref:Ma, ref:KNT, ref:AKS, ref:GNR}.
 Furthermore,
using the physics of extended Higgs sectors
we may consider a model where not only
neutrino masses and dark matter
but also the baryon asymmetry of the universe
can be explained simultaneously
in the context of the electroweak baryogenesis~\cite{ref:AKS}.
 On the other hand,
LNV has not been discovered,
so that nontrivial scenarios to generate masses of Dirac neutrinos
should also be considered.
 Similarly to the cases for Majorana masses,
Dirac masses can be generated at the tree level%
~\cite{ref:nuTHDM-D, ref:DSeesaw}
as well as the loop level%
~\cite{ref:1loopDirac, Gu:2007ug, Kanemura:2016ixx}
involving the dark matter candidate%
~\cite{Gu:2007ug, Kanemura:2016ixx}.

 It is very important to test these models
by using various kinds of current and future experiments.
 Classification of models into several groups
by some common features enables us to effectively test
neutrino mass generation mechanisms
not in model-by-model but in group-by-group.
 In Refs.~\cite{Kanemura:2015cca, Kanemura:2016ixx},
models of neutrino masses are classified
by focusing on the combinations of
new Yukawa coupling matrices for leptons
as we briefly review in the next section.
 Such Yukawa interactions determine
the flavor structure of the neutrino mass matrix.
 If LFV phenomena
other than neutrino oscillations are observed,
the origin of these phenomena
can be the same as that of the new physics for neutrino masses,
because neutrino oscillations show that
lepton flavor conservation is highly violated
in connection to neutrinos.
 The LFV decays of charged leptons
~($\ell \to \ellp \gamma$
and $\tau \to \overline{\ell}_1 \ell_2 \ell_3$)
and the violation of the universality for
$\ell \to \ellp \nu\overline{\nu}$
are considered in Refs.~\cite{Kanemura:2015cca, Kanemura:2016ixx}
for the test of the groups of models.

 By the discovery of the Higgs boson~\cite{ref:2012Jul}
with the mass $125\,\GeV$,
we obtained new observables
to test models of new physics beyond the standard model~(SM).
 In particular,
Higgs boson couplings can be sensitive to new physics effects.
 For example,
LFV decay of the Higgs boson
can be a clear signature of new physics%
~(see e.g.,\ Refs.~\cite{ref:HiggsLFV, ref:HiggsLFV-typeIII,
Blankenburg:2012ex, Harnik:2012pb}).
 The CMS experiment with
the $19.7\,\fb^{-1}$ integrated luminosity at $8\,\TeV$
gives upper bounds on branching ratios
at the 95\,\% confidence level as
$\BR(h \to e\mu) < 3.5\times 10^{-4}$~\cite{Khachatryan:2016rke},
$\BR(h \to e\tau) < 6.9\times 10^{-3}$~\cite{Khachatryan:2016rke},
and $\BR(h \to \mu\tau) < 1.51\times 10^{-2}$~\cite{Khachatryan:2015kon},
where
$\BR(h \to \ell\ellp)
\equiv \BR(h \to \ell\overline{\ellp}) + \BR(h \to \overline{\ell}\ellp)$.
 The best fit value
$\BR(h \to \mu\tau) = 0.84^{+0.39}_{-0.37}\times 10^{-2}$
at the CMS~\cite{Khachatryan:2015kon}
corresponds to the $2.4\,\sigma$ excess.
 The CMS experiment also gives
the best fit value
$\BR(h \to \mu\tau) = -0.76^{+0.81}_{-0.84}\times 10^{-2}$
with $2.3\,\fb^{-1}$ at $13\,\TeV$~\cite{CMS:2016qvi}.
 The ATLAS experiment~\cite{Aad:2016blu}
with $20.3\,\fb^{-1}$ at $8\,\TeV$
obtained upper bounds~(best fit values) as
$\BR(h \to e\tau) < 1.04\times 10^{-2}$%
~($-0.34^{+0.64}_{-0.66}\times 10^{-2}$)
and $\BR(h \to \mu\tau) < 1.43\times 10^{-2}$%
~($0.53^{+0.51}_{-0.51}\times 10^{-2}$).
 See e.g.,\
Refs.~\cite{ref:forCMSexcess-typeIII, ref:forCMSexcess-flavor,
ref:forCMSexcess-radiative, Arganda:2015naa, Herrero-Garcia:2016uab}
for the works to explain the excess at the CMS\@.
 It is expected that
Higgs boson couplings are measured
as precisely as possible
at current and future collider experiments.
 For $\BR(h \to \mu\tau)$,
expected sensitivities are ${\mathcal O}(10^{-4})$
at the LHC~\cite{Han:2000jz}
and the ILC~\cite{Kanemura:2004cn}.
 Even if the excess for $h \to \mu\tau$ at the CMS is not confirmed,
there can be other signal for $h \to \ell\ellp$ in the future.

 In this letter,
we discuss impact of future discoveries of $h \to \ell\ellp$
on the mechanisms to generate neutrino masses.
 Since the Higgs sector is extended in many models for neutrino masses,
such models can naturally connect Higgs physics to LFV phenomena.
 By utilizing systematic analyses
in Refs.~\cite{Kanemura:2015cca, Kanemura:2016ixx}
for mechanisms of neutrino masses,
the simple models for Majorana neutrino masses
cannot be compatible with $h \to \ell\ellp$ signals
because of constraints from $\ell \to \ellp\gamma$,
for which there are no degrees of freedom
for cancellation in these models.
 However,
we find that
some simple models for masses of Dirac neutrinos
can be consistent with $h \to \ell\ellp$ signals
with possible suppression of $\ell \to \ellp\gamma$
by cancellation.
 Namely,
if $h \to \ell\ellp$ is observed,
the observation might indicate that
neutrinos are not Majorana particles
but Dirac particles with lepton number conservation.

 Section~\ref{sec:classify} is devoted to
a brief review of Refs.~\cite{Kanemura:2015cca, Kanemura:2016ixx},
where models of neutrino masses are
systematically classified into some
"Mechanisms"
according to combinations of new Yukawa coupling matrices with leptons.
 In Section~\ref{sec:HiggsLFV},
we discuss LFV decays of the Higgs boson
for simple models in these Mechanisms.
 Conclusions are given in Section~\ref{sec:concl}.




\section{Classification of Models for Generating Neutrino Mass}
\label{sec:classify}


 In Ref.~\cite{Kanemura:2015cca},
all possible Yukawa interactions
between leptons and new scalar fields
are taken into account for
mechanisms to generate Majorana neutrino masses.
 By focusing only on the combinations of such Yukawa coupling matrices
which are the origin of the flavor structure of the neutrino mass matrix,
we can efficiently classify the models
without specifying details of the models,
such as the concrete shape of the scalar potential,
sizes of new coupling constants, and so on.

 In the analyses in Ref.~\cite{Kanemura:2015cca},
the following simplifications are taken:

\noindent
i) No colored scalars~(e.g., leptoquarks) are introduced
in order to concentrate on the lepton sector.

\noindent
ii) Scalar fields do not have flavors
in order to avoid complication.
Therefore,
flavor symmetries and the supersymmetry are not introduced.

\noindent
iii) Each of quarks and leptons
does not interact with two or more $\SU(2)_L$-doublet Higgs fields.
 Then, the flavor changing neutral current~(FCNC) interactions
for quarks and charged leptons are absent at the tree level.
 This can be achieved by using the softly-broken $Z_2$ symmetry,
which is often the case for two Higgs doublet models%
~\cite{Barger:1989fj, Grossman:1994jb, ref:THDM-Akeroyd, Aoki:2009ha}.

\noindent
iv) For Majorana neutrino masses,
only $\psi_R^0$ are introduced as fermions,
which is a singlet under the SM gauge group
with the odd parity for the unbroken $Z_2$ symmetry.
 Therefore,
$(\psi_R^0)^c$ are not mixed with $\nu_L^{}$,
which are $Z_2$-even. 
 The type-I and type-III seesaw mechanisms,
where new fermions are mixed with $\nu_L^{}$,
are not included in the analyses
because new physics effects of them at the low energy
are highly suppressed by large masses of new fermions.
 Of course,
right handed neutrinos $\nu_R$ are also introduced
for analyses of masses of Dirac neutrinos
in Ref.~\cite{Kanemura:2016ixx}.

\noindent
v) Three tiny neutrino masses
are generated by a diagram.
 Introduced scalar fields
are only the ones that are necessary for the diagram.

 It was found
that only four combinations%
~(the Mechanisms-M1 -- M4 in Table~\ref{tab:M})
of new Yukawa interactions%
~(or equivalently new scalar fields)
can generate Majorana neutrino masses.
 Although another combination exists in principle,
which corresponds the case in a simplified version of the Zee model
such that there is no FCNC at the tree-level%
~\cite{Zee:1980ai,Wolfenstein:1980sy},
the flavor structure of the neutrino mass matrix
has already been excluded by neutrino oscillation data~\cite{He:2003ih}.
 There appear additional four combinations%
~(the Mechanisms-M5 -- M8 in Table~\ref{tab:M})
if singlet fermions $\psi_R^0$ and
additional scalar fields for Yukawa interactions
between $\psi_R^0$ and leptons
are introduced with the odd parity under an unbroken $Z_2$ symmetry.
 Such $Z_2$-odd particles can provide the dark matter candidate.
 In Ref.~\cite{Kanemura:2015cca},
it was also found
that these eight Mechanisms
can be further classified into only three
"Groups"
according to the combination of
new interactions between two leptons,
where $\psi_R^0$ are integrated out.
 These Groups
can be tested
by measurements of the absolute neutrino mass,
the neutrinoless double beta decay
and by $\tau \to \overline{\ell}_1 \ell_2 \ell_3$.
 Predictions in these Groups
are not applicable to the type-I~(and III) seesaw scenario
because of the absence of new scalar particles.
 Notice that representations of new scalar fields
associated with the new interaction between two leptons
are hidden by the classification into Groups,
e.g., the interaction between two $\ell_R$
can be accompanied with
a doubly-charged scalar, two singly-charged scalars
or some other scalar fields.
 In this letter,
we rely on the classification into 
not Groups but Mechanisms
in order to discuss the chiral structure for $\ell \to \ellp\gamma$,
which requires representations of scalar fields to be fixed.


 In Table~\ref{tab:M},
we show the combinations of new scalar fields
that can generate Majorana neutrino masses.
 Scalar fields $s_L^+$, $s^{++}$, and $s_2^+$
are all singlet under $\SU(2)_L$.
 Fields $s_L^+$ and $s_2^+$ have hypercharge $Y=1$
while $s^{++}$ has $Y=2$.
 The second $\SU(2)_L$-doublet field $\Phi_2$ has $Y=1/2$.
 In order to avoid the FCNC at the tree level,
each of right-handed quarks and leptons
has the Yukawa interaction with only an $\SU(2)_L$-doublet Higgs field
by implicitly introducing softly-broken $Z_2$ symmetries%
~\cite{Barger:1989fj, Grossman:1994jb, ref:THDM-Akeroyd, Aoki:2009ha}.
 In this letter,
we take such that $\ell_R$ couples with $\Phi_2$
without loss of generality.
 Another $\SU(2)_L$-doublet field
$\eta = (\eta^+, \eta^0)^T$ with $Y=1/2$
as well as $s_2^+$ and gauge singlet fermions $\psi_R^0$
are odd under the unbroken $Z_2$ symmetry.
 The $\SU(2)_L$-triplet field%
\footnote{%
 The FCNC for $\nu_L^{}$ via $\Delta^0$ is acceptable.
}
with $Y=1$ is denoted by $\Delta$.
 Simple realizations of these Mechanisms
correspond to the models in references in the last column,
where scalar lines for these Mechanisms
are explicitly closed by using appropriate scalar interactions.


 Similarly,
the classification of models to generate masses of Dirac neutrinos
is achieved in Ref.~\cite{Kanemura:2016ixx},
where $\nu_R^{}$ are introduced
with the lepton number conservation.
 In order to forbid the Yukawa interaction
of neutrinos with the SM Higgs doublet field,
the softly-broken $Z_2$ symmetry~(denoted as $Z_2^\prime$)
is also introduced such that
$\nu_R^{}$ has the odd parity
while fields exist in the SM have the even parity.
 It was shown that
Dirac neutrino masses can be generated
by seven combinations of new Yukawa coupling matrices%
~(the Mechanisms-D1 -- D7 in Table~\ref{tab:D}).
 If we introduce $Z_2$-odd fields~(e.g., $\psi_R^0$)
similarly to the cases for Majorana neutrino masses,
additional eleven combinations
~(the Mechanisms-D8 -- D18 in Table~\ref{tab:D})
can generate Dirac neutrino masses.
 These eighteen Mechanisms
to generate Dirac neutrino masses
can be further classified into seven Groups
according to the combination of
new interactions between two leptons,
where $\psi_R^0$ are integrated out.
 Some of these Groups
can be tested
by measurements of the absolute neutrino mass
and $\tau \to \ellp \nu\overline{\nu}$~\cite{Kanemura:2016ixx}.


 The combinations of new scalar fields
for masses of Dirac neutrinos
are listed in Table~\ref{tab:D}.
 Scalar fields $s^0$, $s_R^+$, and $s_2^0$
are all singlet%
\footnote{%
 The FCNC for $\nu_R^{}$ via $s^0$ is acceptable.
}
under $\SU(2)_L$.
 Hypercharges of $s^0$ and $s_2^0$ are zero,
and $s_R^+$ has $Y=1$.
 Since $s_R^+$ and $s_2^0$ are $Z_2^\prime$-odd fields,
they can couple to a $\nu_R$.
 This property is the difference of $s_R^+$ from $s_L^+$.
 The $Z_2^\prime$-odd field $\Phi_\nu$
is an $\SU(2)_L$-doublet field with $Y=1/2$,
which has Yukawa interaction only with $\nu_R$.
 For the cases of Dirac neutrino masses,
conserving lepton numbers
are assigned to these new scalar fields
as shown in Table~\ref{tab:D}.
 Singlet fermions $\psi_R^0$ do not have the lepton number,
and then they can have Majorana mass terms without the LNV\@.
 Due to these assignments of conserving lepton numbers,
an unbroken $Z_2$ symmetry appears automatically.

\begin{table}[t]
\begin{tabular}
{c||
>{\centering\arraybackslash}p{8mm}|
>{\centering\arraybackslash}p{8mm}|
>{\centering\arraybackslash}p{8mm}|
>{\centering\arraybackslash}p{8mm}||
>{\centering\arraybackslash}p{8mm}|
>{\centering\arraybackslash}p{8mm}||
>{\centering\arraybackslash}p{8mm}|
>{\centering\arraybackslash}p{8mm}||
c}
{}
 & \multicolumn{6}{c||}{ Scalar with leptonic Yukawa int. }
 & \multicolumn{2}{c||}{  }
 &
\\
\cline{2-7}
{}
 & \multicolumn{4}{c||}{}
 & \multicolumn{2}{c||}{$Z_2$-odd}
 & \multicolumn{2}{c||}{ $\ell \to \ellp \gamma$ }
 &
\\
\cline{2-7}\cline{8-9}
 & $s_L^+$
 & $s^{++}$
 & $\Phi_2$
 & $\Delta$
 & $\red{s_2^+}$
 & $\red{\eta}$
 & $\ellp_L$
 & $\ellp_R$
 &
\\
\cline{1-7}
 $\SU(2)_L$
 & $\underline{\bf 1}$
 & $\underline{\bf 1}$
 & $\underline{\bf 2}$
 & $\underline{\bf 3}$
 & $\underline{\bf 1}$
 & $\underline{\bf 2}$
 & 
 & 
 &
\\
\cline{1-7}
 $\U(1)_Y$
 & $1$
 & $2$
 & $1/2$
 & $1$
 & $1$
 & $1/2$
 & 
 & 
 & 
\\
\cline{1-7}
 Unbroken $Z_2$
 & $+$
 & $+$
 & $+$
 & $+$
 & $-$
 & $-$
 & 
 & 
 & Simple models
\\
\hline\hline
%
%
M1
 & $\checkmark$
 & $\checkmark$
 &
 &
 &
 &
 & $\checkmark$
 & $\checkmark$
 & \cite{Zee:1985id, Babu:1988ki}
\\
\hline
%
%
M2
 &
 & $\checkmark$
 & $\checkmark$
 &
 &
 &
 &
 & $\checkmark$
 & \cite{Cheng:1980qt, Kanemura:2015cca}
\\
\hline
%
%
M3
 &
 & $\checkmark$
 &
 &
 &
 &
 &
 & $\checkmark$
 & \cite{ref:GNR}
\\
\hline
%
%
M4
 &
 &
 &
 & $\checkmark$
 &
 &
 & $\checkmark$
 &
 & \cite{Cheng:1980qt, ref:HTM}
\\
\hline
%
%
M5
 & $\checkmark$
 &
 &
 &
 & $\checkmark$
 &
 & $\checkmark$
 & $\checkmark$
 & \cite{ref:KNT}
\\
\hline
%
%
M6
 &
 &
 & $\checkmark$
 &
 & $\checkmark$
 &
 &
 & $\checkmark$
 & \cite{ref:AKS}
\\
\hline
%
%
M7
 &
 &
 &
 &
 & $\checkmark$
 &
 &
 & $\checkmark$
 & This letter${}^4$
\\
\hline
%
%
M8
 &
 &
 &
 &
 &
 & $\checkmark$
 & $\checkmark$
 &
 & \cite{ref:Ma}
\end{tabular}
\caption{
 It shows which scalar fields
are introduced in the Mechanisms-M1 -- M8,
which generate Majorana neutrino masses.
 A check-mark means that
the Mechanism includes the scalar field.
 Columns of $\ellp_L$ and $\ellp_R$ show
the chirality of $\ellp$ of $\ell \to \ellp \gamma$ in each Mechanism.
}
\label{tab:M}
\end{table}

\begin{table}[t!h]
\begin{tabular}
{c||
>{\centering\arraybackslash}p{7mm}|
>{\centering\arraybackslash}p{7mm}|
>{\centering\arraybackslash}p{7mm}|
>{\centering\arraybackslash}p{7mm}|
>{\centering\arraybackslash}p{7mm}|
>{\centering\arraybackslash}p{7mm}|
>{\centering\arraybackslash}p{7mm}||
>{\centering\arraybackslash}p{7mm}|
>{\centering\arraybackslash}p{7mm}|
>{\centering\arraybackslash}p{7mm}||
>{\centering\arraybackslash}p{7mm}|
>{\centering\arraybackslash}p{7mm}||c}
{}
 & \multicolumn{10}{c||}{ Scalar with leptonic Yukawa int. }
 & \multicolumn{2}{c||}{  }
 &
\\
\cline{2-11}
{}
 & \multicolumn{7}{c||}{}
 & \multicolumn{3}{c||}{$Z_2$-odd}
 & \multicolumn{2}{c||}{ $\ell \to \ellp \gamma$ }
 &
\\
\cline{2-11}\cline{12-13}
 & $s^0$
 & $s_L^+$
 & $s_R^+$
 & $s^{++}$
 & $\Phi_\nu$
 & $\Phi_2$
 & $\Delta$
 & $\red{s_2^0}$
 & $\red{s_2^+}$
 & $\red{\eta}$
 & $\ellp_L$
 & $\ellp_R$
 &
\\
\cline{1-11}
 $\SU(2)_L$
 & $\underline{\bf 1}$
 & $\underline{\bf 1}$
 & $\underline{\bf 1}$
 & $\underline{\bf 1}$
 & $\underline{\bf 2}$
 & $\underline{\bf 2}$
 & $\underline{\bf 3}$
 & $\underline{\bf 1}$
 & $\underline{\bf 1}$
 & $\underline{\bf 2}$
 & 
 & 
 &
\\
\cline{1-11}
 $\U(1)_Y$
 & $0$
 & $1$
 & $1$
 & $2$
 & $1/2$
 & $1/2$
 & $1$
 & $0$
 & $1$
 & $1/2$
 & 
 & 
 &
\\
\cline{1-11}
 Lepton number
 & $-2$
 & $-2$
 & $-2$
 & $-2$
 & $0$
 & $0$
 & $-2$
 & $-1$
 & $-1$
 & $-1$
 & 
 & 
 &
\\
\cline{1-11}
 $Z_2^\prime$
 & $+$
 & $+$
 & $-$
 & $+$
 & $-$
 & $+$
 & $+$
 & $-$
 & $+$
 & $+$
 & 
 & 
 & Simple models
\\
\hline\hline
%
%
D1
 &
 & $\checkmark$
 & $\checkmark$
 &
 &
 &
 &
 &
 &
 &
 & $\checkmark$
 & $\checkmark$
 & \cite{ref:1loopDirac}
\\
\hline
%
%
D2
 &
 &
 & $\checkmark$
 &
 &
 &
 & $\checkmark$
 &
 &
 &
 & $\checkmark$
 & $\checkmark$
 & \cite{Kanemura:2016ixx}
\\
\hline
%
%
D3
 &
 &
 & $\checkmark$
 & $\checkmark$
 &
 & $\checkmark$
 &
 &
 &
 &
 &
 & $\dcheckmark$
 & \cite{Kanemura:2016ixx}
\\
\hline
%
%
D4
 &
 &
 & $\checkmark$
 & $\checkmark$
 &
 &
 &
 &
 &
 &
 &
 & $\dcheckmark$
 & \cite{Kanemura:2016ixx}
\\
\hline
%
%
D5
 & $\checkmark$
 &
 & $\checkmark$
 &
 &
 & $\checkmark$
 &
 &
 &
 &
 &
 & $\checkmark$
 & \cite{Kanemura:2016ixx}
\\
\hline
%
%
D6
 & $\checkmark$
 &
 & $\checkmark$
 &
 &
 &
 &
 &
 &
 &
 &
 & $\checkmark$
 & \cite{Kanemura:2016ixx}
\\
\hline
%
%
D7
 &
 &
 &
 &
 & $\checkmark$
 &
 &
 &
 &
 &
 & $\checkmark$
 &
 & \cite{ref:nuTHDM-D}
\\
\hline\hline
%
%
D8
 &
 & $\checkmark$
 &
 &
 &
 &
 &
 & $\checkmark$
 & $\checkmark$
 &
 & $\checkmark$
 & $\checkmark$
 & \cite{Kanemura:2016ixx}
\\
\hline
%
%
D9
 &
 &
 &
 &
 &
 &
 & $\checkmark$
 & $\checkmark$
 & $\checkmark$
 &
 & $\checkmark$
 & $\checkmark$
 & \cite{Kanemura:2016ixx}
\\
\hline
%
%
D10
 &
 &
 & $\checkmark$
 &
 &
 &
 &
 &
 &
 & $\checkmark$
 & $\checkmark$
 & $\checkmark$
 & \cite{Kanemura:2016ixx}
\\
\hline
%
%
D11
 &
 &
 & $\checkmark$
 &
 &
 & $\checkmark$
 &
 &
 & $\checkmark$
 &
 &
 & $\dcheckmark$
 & \cite{Kanemura:2016ixx}
\\
\hline
%
%
D12
 &
 &
 & $\checkmark$
 &
 &
 &
 &
 &
 & $\checkmark$
 &
 &
 & $\dcheckmark$
 & \cite{Kanemura:2016ixx}
\\
\hline
%
%
D13
 &
 &
 & $\checkmark$
 &
 &
 & $\checkmark$
 &
 & $\checkmark$
 &
 &
 &
 & $\checkmark$
 & \cite{Kanemura:2016ixx}
\\
\hline
%
%
D14
 &
 &
 & $\checkmark$
 &
 &
 &
 &
 & $\checkmark$
 &
 &
 &
 & $\checkmark$
 & \cite{Kanemura:2016ixx}
\\
\hline
%
%
D15
 &
 &
 &
 &
 &
 & $\checkmark$
 &
 & $\checkmark$
 & $\checkmark$
 &
 &
 & $\checkmark$
 & \cite{Kanemura:2016ixx}
\\
\hline
%
%
D16
 &
 &
 &
 &
 &
 &
 &
 & $\checkmark$
 & $\checkmark$
 &
 &
 & $\checkmark$
 & \cite{Kanemura:2016ixx}
\\
\hline
%
%
D17
 &
 &
 & $\checkmark$
 &
 &
 &
 &
 &
 & $\checkmark$
 & $\checkmark$
 & $\checkmark$
 & $\dcheckmark$
 & \cite{Kanemura:2016ixx}
\\
\hline
%
%
D18
 &
 &
 &
 &
 &
 &
 &
 & $\checkmark$
 &
 & $\checkmark$
 & $\checkmark$
 &
 & \cite{Gu:2007ug}
\end{tabular}
\caption{
 It shows which scalar fields
are introduced in the Mechanisms-D1 -- D18,
which generate Dirac neutrino masses.
 A check-mark means that
the Mechanism includes the scalar field.
 Columns of $\ellp_L$ and $\ellp_R$ show
the chirality of $\ellp$ of $\ell \to \ellp \gamma$ in each Mechanism.
 Two check-marks in a cell for $\ellp_R$ mean
that two scalar fields contribute to $\ell \to \ellp_R \gamma$.
}
\label{tab:D}
\end{table}



\section{Lepton Flavor Violating Higgs Boson Decay}
\label{sec:HiggsLFV}

\begin{figure}[t]
  \begin{center}
   \includegraphics[scale=0.7]{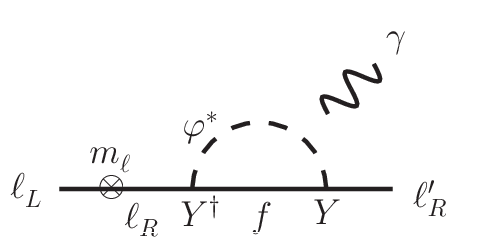}
   \includegraphics[scale=0.7]{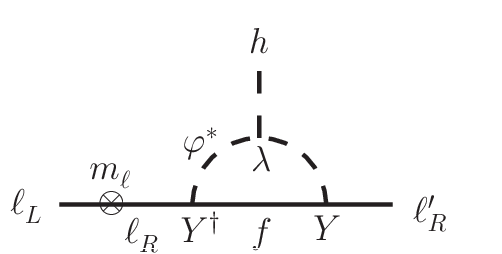}
   \vspace*{-5mm}
   \caption{
    Diagrams for $\ell \to \ellp\gamma$~(left)
    and $h \to \ell\ellp$~(right).
   }
   \label{fig:LFV}
  \end{center}
\end{figure}

 In this section,
we discuss $h \to \ell\ellp$
in order to clarify the impact of future discovery of the decay
on the Mechanisms in Tables~\ref{tab:M} and \ref{tab:D}.
 First of all,
let us take only a new Yukawa interaction
$Y_{a \ell}\, \overline{f_a}\, \ell_X\, \varphi$
between a charged lepton $\ell$
and a charged%
\footnote{
 We assume that there is no FCNC
for quarks and charged leptons at the tree level.
}%
%
\footnotetext{
 Scalar lines for the Mechanism-M7 can be closed
by introducing a real $\SU(2)_L$-triplet scalar $\Delta_2$~($Z_2$-odd)
via $\Phi^T \epsilon \Delta_2 \Phi s_2^-$.
}
%
scalar $\varphi$,
where $X = L, R$ denote chirality of $\ell$.
 The particle $f$ is a certain fermion.
 For example,
the Zee-Babu model~\cite{Zee:1985id, Babu:1988ki} of the Mechanism-M1
has the interaction with $f = (\ell_R)^c$,
$\ell_X=\ell_R$, and $\varphi=s^{++}$;
 for the Ma model~\cite{ref:Ma} of the Mechanism-M8,
$f = \psi_R^0$, $\ell_X=\ell_L$, and $\varphi=\eta^+$.
 This interaction causes $\ell \to \ell_X^\prime \gamma$
with the diagram in Fig.~\ref{fig:LFV}~(left),
whose branching ratio is given by
\begin{eqnarray}
\BR(\ell \to \ellp_X \gamma)
\simeq
 \begin{cases}
  \displaystyle
   \frac{ \alpha \pi^4 }{ 3 (16\pi^2)^2 G_F^2 }
   \frac{
         (2-3 Q_\varphi)^2
         \left| S^2 ( Y^\dagger Y )_{\ell\ellp} \right|^2
        }
        { m_\varphi^4 }
   \BR(\ell \to e \nu_\ell^{} \overline{\nu_e^{}})
    & ( m_f \ll m_\varphi )
\\[3mm]
  \displaystyle
   \frac{ \alpha \pi^4 }{ 3 (16\pi^2)^2 G_F^2 }
   \frac{
         (1-3 Q_\varphi)^2
         \left| S^2 ( Y^\dagger Y )_{\ell\ellp} \right|^2
        }
        { m_f^4 }
   \BR(\ell \to e \nu_\ell^{} \overline{\nu_e^{}})
    & ( m_f \gg m_\varphi )
\\
 \end{cases} ,
\label{eq:ltolpgam}
\end{eqnarray}
where $G_F$ is the Fermi constant,
$\alpha$ is the fine structure constant,
and $Q_\varphi$ is the electric charge of $\varphi$.
 The electric charge of $f$ is $Q_\varphi - 1$.
 Masses of $\varphi$ and $f_a$ are denoted as
$m_\varphi$ and $m_f$~(assumed to be common for $f_a$),
respectively.
 The factor $S$ is taken to be $2$ for the case
where the Yukawa matrix $Y$ is symmetric or antisymmetric,
and $1$ for the other cases.

 The new Yukawa interaction used above also gives
the lepton flavor violating decay of the Higgs boson $h$
at the one-loop level as shown in Fig.~\ref{fig:LFV}~(right).
 The decay branching ratio
$\BR(h\to \ell\ell^\prime)$%
~($\equiv \BR(h \to \ell\overline{\ellp})
 + \BR(h \to \overline{\ell}\ellp)$),
where $\ell\neq \ellp$ and $m_\ell > m_{\ellp}$,
can be calculated as
\begin{eqnarray}
\BR( h \to \ell\ellp )
&\simeq&
 \begin{cases}
  \displaystyle
  \frac{ v^2 m_h^{} }
       {128 \pi (16\pi^2)^2 \Gamma_\text{tot} }\,
  \frac{
        \lambda^2 m_\ell^2
        \left| S^2 (Y^\dagger Y)_{\ell\ellp} \right|^2
       }
       { m_\varphi^4 }
    & ( m_f \ll m_\varphi )
\\[3mm]
  \displaystyle
  \frac{ v^2 m_h^{} }
       {128 \pi (16\pi^2)^2 \Gamma_\text{tot} }
  \frac{
        \lambda^2 m_\ell^2
        \left| S^2 (Y^\dagger Y)_{\ell\ellp} \right|^2
       }
       { m_f^4 }
  \left(
   3 - \ln\frac{m_\psi^2}{m_\varphi^2}
  \right)^2
    & ( m_f \gg m_\varphi )
 \end{cases} ,
\label{eq:htollp}
\end{eqnarray}
where $\lambda$ is the coupling constant
of the interaction $\lambda v h |\varphi|^2$
with the vacuum expectation value $v$~($= 246\,\GeV$).
 The Higgs boson mass is denoted by $m_h$~($=125\,\GeV$),
and $\Gamma_\text{tot}$ stands for
the total width of the Higgs boson~\cite{ref:CERN-YRP-BR}.
 With the ratio of eqs.~\eqref{eq:ltolpgam} and \eqref{eq:htollp},
it is clear that
magnitudes of $\BR(h \to \ell\ellp)$ and $\BR(\ell \to \ellp \gamma)$
are similar to each other
except for the cases with $Q_\varphi = 2/3$%
~(see e.g., Ref.~\cite{Davidson:1993qk} for leptoquarks)
and $1/3$.
 Under the constraint from the current bounds
$\BR( \mu \to e \gamma) < 4.2 \times 10^{-13}$~\cite{TheMEG:2016wtm}
and $\BR( \tau \to \ellp \gamma) \lesssim 10^{-8}$~\cite{Aubert:2009ag},
$\BR(h \to \ell\ellp)$ is too small to be observed
if it is radiatively produced.
 If $\BR(h \to \ell\ellp)$ is observed,
such a simple model is excluded.
 Then,
we might take FCNC at the tree level
in order to explain the signal%
~\cite{ref:HiggsLFV-typeIII, ref:forCMSexcess-typeIII}
or take some extension
to suppress $\ell \to \ellp \gamma$ by cancellation%
~(see, e.g.\ Ref.~\cite{ref:forCMSexcess-radiative}
for the cancellation).

 Each of the Mechanisms
listed in Tables~\ref{tab:M} and \ref{tab:D}
has new Yukawa interactions with charged leptons,
which can produce both $\ell \to \ellp \gamma$ and $h \to \ell\ellp$.
 According to the discussion in the previous paragraph,
Mechanisms for which there is only a check-mark
in columns of $\ell \to \ellp\gamma$
will be excluded if $h \to \ell\ellp$ is really observed.
 Although the Mechanisms-M1, M5, D1, D2, D8, D9 and D10
have two kinds of new Yukawa interactions with charged leptons,
their effects to $\ell \to \ellp \gamma$
cannot be cancelled with each other
because of different chiralities of charged leptons in these interactions.
 For example,
$s^+$ in the Zee-Babu model~\cite{Zee:1985id, Babu:1988ki} of the Mechanism-M1
gives $\ell \to \ellp_L \gamma$
via
$(Y^s_A)_{\ell\ellp}
\Bigl[ \overline{L_\ell}\,\epsilon\, L_{\ellp}^\ast\, s_L^- \Bigr]$
while $s^{++}$ in the model
does $\ell \to \ellp_R \gamma$
via
$(Y^s_S)_{\ell\ellp}
\Bigl[ \overline{(\ell_R)^c}\, \ellp_R\, s^{++} \Bigr]$.
 Even in the type-I and III seesaw scenarios,
$\BR(h \to \ell\ellp)/\BR(\ell \to \ellp\gamma)$
is not enhanced.
 This means that
all Mechanisms for Majorana neutrino masses in Table~\ref{tab:M}
as well as the type-I and III seesaw scenarios
are not suitable as low-energy effective theories
if $h \to \ell\ellp$ is observed.
 Exclusion of some specific models for neutrino masses
are shown in Ref.~\cite{Herrero-Garcia:2016uab}.
 Our statement covers the wider class
of models to generate neutrino masses
by virtue of systematic classification of the models.

\begin{figure}[t]
  \begin{center}
   \includegraphics[scale=0.7]{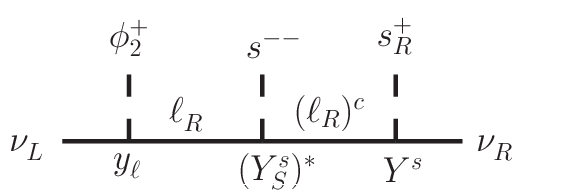}
   \vspace*{-5mm}
   \caption{
    The diagram of the Mechanism-D3.
    (Taken from Ref.~\cite{Kanemura:2016ixx}.)
   }
   \label{fig:D3}
  \end{center}
\end{figure}

\begin{figure}[t]
  \begin{center}
   \includegraphics[scale=0.7]{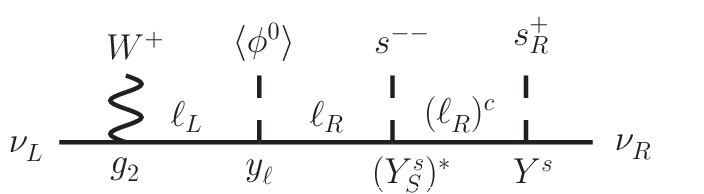}
   \vspace*{-5mm}
   \caption{
    The diagram of the Mechanism-D4.
    (Taken from Ref.~\cite{Kanemura:2016ixx}.)
   }
   \label{fig:D4}
  \end{center}
\end{figure}

\begin{figure}[t]
  \begin{center}
   \includegraphics[scale=0.7]{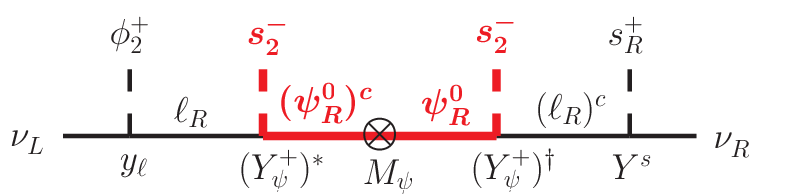}
   \vspace*{-5mm}
   \caption{
    The diagram of the Mechanism-D11.
    (Taken from Ref.~\cite{Kanemura:2016ixx}.)
   }
   \label{fig:D11}
  \end{center}
\end{figure}

\begin{figure}[t]
  \begin{center}
   \includegraphics[scale=0.7]{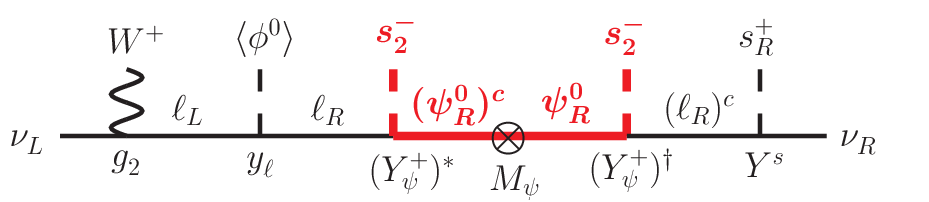}
   \vspace*{-5mm}
   \caption{
    The diagram of the Mechanism-D12.
    (Taken from Ref.~\cite{Kanemura:2016ixx}
     correcting a typo as $Y_S^0 \to Y^s$.)
   }
   \label{fig:D12}
  \end{center}
\end{figure}

\begin{figure}[t]
  \begin{center}
   \includegraphics[scale=0.7]{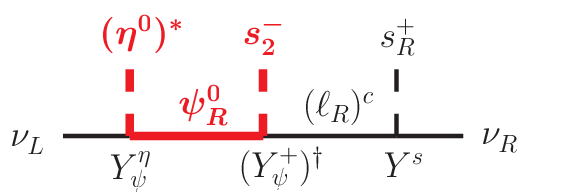}
   \vspace*{-5mm}
   \caption{
    The diagram of the Mechanism-D17.
    (Taken from Ref.~\cite{Kanemura:2016ixx}.)
   }
   \label{fig:D17}
  \end{center}
\end{figure}

 On the other hand,
it is found that
some Mechanisms for Dirac neutrino masses in Table~\ref{tab:D}
can be compatible with the observation of $h \to \ell\ellp$.
 In both of the Mechanisms-D3 and D4,
$s_R^+$ and $s^{++}$ interact with $\ell_R$
via
Yukawa interactions
$(Y^s)_{\ell i}
\Bigl[ \overline{(\ell_R)^c}\, \nu_{iR}^{}\, s_R^+ \Bigr]$
and
$(Y^s_S)_{\ell\ellp}
\Bigl[ \overline{(\ell_R^{})^c}\, \ell^\prime_R\, s^{++} \Bigr]$,
respectively.
 Figure~\ref{fig:D3} for the Mechanism-D3
and Figure~\ref{fig:D4} for the Mechanism-D4
show how $\nu_L^{}$ is connected to $\nu_R^{}$
in order to generate the Dirac neutrino mass,
where $y_\ell = \sqrt{2}\,m_\ell/v$,
and $g_2$ is the $\SU(2)_L$ gauge coupling constant.
 Contributions of these scalars
to $\ell \to \ellp_R \gamma$
can be destructive such as
\begin{eqnarray}
\BR( \ell \to \ellp\gamma )
\propto
 \left|
  (-1)
  \frac{ (Y^{s\dagger} Y^s)_{\ell\ellp} }{ m_{s^+_R}^2 }
  +
  (-16)
  \frac{ (Y^{s\dagger}_S Y^s_S)_{\ell\ellp} }{ m_{s^{++}}^2 }
 \right|^2
\ll
 \left|
  \frac{ (Y^{s\dagger} Y^s)_{\ell\ellp} }{ m_{s^+_R}^2 }
 \right|^2 ,
\end{eqnarray}
where $m_{s^+_R}^{}$ and $m_{s^{++}}^{}$
are masses of $s^+_R$ and $s^{++}$, respectively.
 For example,
since $\BR( h \to \mu\tau ) \sim 10^{-3}$
naively corresponds to $\BR( \tau \to \mu\gamma ) \sim 10^{-2}$
for $m_f \ll m_\varphi$
with $\lambda^2/(2 - 3Q_\varphi)^2 \sim 1$,
the $10^{-3}$ tuning of two amplitudes
is required for the cancellation to satisfy
$\BR( \tau \to \mu \gamma ) \lesssim 10^{-8}$.
 Even in such cases,
contributions of two scalar fields to $h \to \ell\ellp$
can be constructive by utilizing coupling constants for interactions
$\lambda_{h s^+} v h |s_R^+|^2$
and $\lambda_{h s^{++}} v h |s^{++}|^2$ such as
\begin{eqnarray}
\BR( h \to \ell\ellp )
\propto
 \left|
  \lambda_{ h s^+ }
  \frac{ (Y^{s\dagger} Y^s)_{\ell\ellp} }{ m_{s^+_R}^2 }
  +
  4
  \lambda_{ h s^{++} }
  \frac{ (Y^{s\dagger}_S Y^s_S)_{\ell\ellp} }{ m_{s^{++}}^2 }
 \right|^2
\sim
 \left|
  \lambda_{ h s^+ }
  \frac{ (Y^{s\dagger} Y^s)_{\ell\ellp} }{ m_{s^+_R}^2 }
 \right|^2 ,
\end{eqnarray}
where $\lambda_{h s^+}$ and $\lambda_{h s^{++}}$
should have the opposite sign.
 Notice that these interactions of scalars are
not used to close scalar lines of
the diagrams~(Figs.~\ref{fig:D3} and \ref{fig:D4})
for the neutrino mass generation,
and then they are free from constraints
from neutrino oscillation experiments.
 Some explicit examples to close the scalar lines
are shown in Ref.~\cite{Kanemura:2016ixx}.
 This is also the case for
the Mechanisms-D11, D12, and D17%
\footnote{
 In the Mechanism-D17,
a diagram with the chirality flip via the mass of $\psi_R^0$
seems to contribute to $h \to \ell\ellp$
by using the $h \eta^+ s_2^-$ interaction.
 However,
the contribution is understood as
a dimension-4 operator,
and such a contribution disappears
by the diagonalization of charged lepton mass matrix
at the loop level%
~(see e.g., Ref.~\cite{Harnik:2012pb}).
},
in which $s_R^+$ and $s_2^+$ interact with $\ell_R$
via
$(Y^s)_{\ell i}
\Bigl[ \overline{(\ell_R)^c}\, \nu_{iR}^{}\, s_R^+ \Bigr]$
and
$(Y_\psi^+)_{\ell i}
\Bigl[ \overline{(\ell_R^{})^c}\, \psi_{iR}^0\, s_2^+ \Bigr]$,
respectively.
 Dirac neutrino masses
are generated by connecting $\nu_L^{}$ to $\nu_R^{}$
as shown in Figs.~\ref{fig:D11}, \ref{fig:D12}, and \ref{fig:D17},
where the Majorana mass term
$(1/2) M_\psi \Bigl[ \overline{(\psi_R^0)^c} \psi_R^0 \Bigr]$
and the Yukawa interaction
$(Y_\psi^\eta)_{\ell i}
\Bigl[ \overline{L_\ell} \epsilon \eta^\ast \psi_{iR}^0 \Bigr]$
are utilized.
 Therefore,
these Mechanisms of the Dirac neutrino mass would be preferred
when $h \to \ell\ellp$ is observed.

 As discussed in Ref.~\cite{Kanemura:2016ixx},
the Mechanisms-D3, D4, D11, and D12
can be classified further into a Group
that gives the Dirac neutrino mass matrix
$m_\text{D} \propto y_\ell X_{SR}^\ast Y^s$,
where the symmetric matrix $X_{SR}$
corresponds to the (effective) interaction between $\ell_R$ and $(\ell_R)^c$.
 The case with $X_{SR} = Y_S^s$
gives the Mechanisms-D3 and D4,
and the case with $X_{SR} = (Y_\psi^+)^\ast M_\psi (Y_\psi^+)^\dagger$
does the Mechanisms-D11 and D12.
 Multiplying $y_\ell^{-1}$ from the left-hand side,
it is expected that
some of the new Yukawa interactions prefer to couple to the electron
because of the hierarchical structure of $y_\ell^{-1}$.
 Therefore,
fine-tuning to suppress $\mu \to e\gamma$
might be required.
 Notice that
the effective interaction $h\overline{\mu} e$
should also be suppressed
in order to avoid its contribution to
$\mu \to e\gamma$ at the loop level
involving $h$ in the loop~\cite{Blankenburg:2012ex, Harnik:2012pb}.
 On the other hand,
the Mechanism-D17 is not suffered from
such an enhanced interaction with the electron
because the Mechanism gives
$m_\text{D}^{} \propto Y_\psi^\eta (Y_\psi^+)^\dagger Y^s$,
in which $y_\ell$ is not involved.

 In addition to $h \to \ell\ellp$,
a discovery of the second scalar
will make it possible to narrow down the Mechanisms.
 If the CP-odd Higgs boson $A^0$ is discovered,
the Mechanisms-D3 and D11 in which $\Phi_2$ is involved
are selected as candidates for viable Mechanisms.
 Notice that the neutral component of $\eta$ in the Mechanism-D17
is a complex scalar (not divided into CP-even and odd ones)
because it has the lepton number.
 Existence of $\SU(2)_L$-doublet $\eta$,
which has no vacuum expectation value,
is characteristic in the Mechanism-D17\@.
 The Mechanisms-D3 and D11 can also be supported
by discovery of a singly-charged scalar ($s_R^-$)
that dominantly decays into $\tau\overline{\nu}$,
similarly to the case of the type-X THDM
with a large $\tan\beta$%
~\cite{Aoki:2009ha, Kanemura:2014bqa}.
 Discovery of a doubly charged scalar
that decays into a pair of same-signed charged leptons%
\footnote{%
 Simple examples to close scalar lines
for the Mechanisms-D12 and D17 are
shown in Ref.~\cite{Kanemura:2016ixx}
by additionally introducing
the $\SU(2)_L$-doublet scalar field with $Y=3/2$.
 However, its doubly-charged component does not decay
into a pair of same-sign charged leptons in the example.
 See also Ref.~\cite{Aoki:2011yk},
where the doublet scalar field with $Y=3/2$
is utilized to generate neutrino masses.
}
indicates the Mechanisms-D3 and D4\@.


 We here give a comment on some exceptions to the discussion above
when $h \to \ell\ellp$ is detected.
%
 First,
some Mechanisms in Tables~\ref{tab:M} and \ref{tab:D}
include the second Higgs doublet field $\Phi_2$,
which can give the FCNC at the tree-level
similarly to the type-III THDM~\cite{Liu:1987ng}
though we assumed the absence of that.
 Then,
$h \to \ell\ellp$ can happen at the tree-level
while $\ell \to \ellp\gamma$ can be suppressed
as a loop-level process.
 If we accept the FCNC at the tree-level
within experimental constraints,
the Zee model can be consistent with the neutrino oscillation data%
~\cite{He:2003ih}.
 The discovery of $A^0 \to \ell\ellp$
would indicate such cases.
 Since radiative mechanisms for $h \to \ell\ellp$
discussed in this letter
rely on the interaction $\lambda |\Phi_1|^2 |\varphi|^2$,
where $\Phi_1$ denotes the SM-like Higgs doublet,
there is no $A^0 \to \ell\ellp$ with these mechanisms.
%
 Second,
there can be a new Yukawa interaction of charged leptons
with a charged scalar whose electric charge is $2/3$ or $1/3$.
 Their contributions to $\ell \to \ellp\gamma$ are
suppressed, e.g., by a factor of $m_\psi^4/m_\varphi^4$.
 Such a Yukawa interaction was not taken into account in our analyses
because we used only Yukawa interactions between two leptons
or between a lepton and a singlet fermion $\psi_R^0$.
 Discovery of new particle associated with
leptons and quarks would indicate such cases.
%
 Third,
models for the neutrino mass can be extended
by introducing copies of scalar fields.
 Then,
we can utilize cancellation of their contributions
to $\ell \to \ellp\gamma$.
 If two kinds of doubly charged scalars are discovered,
such extensions would be indicated.




\section{Conclusions}
\label{sec:concl}

 We have studied the LFV decay of the Higgs boson
in a wide set of models for neutrino masses
where new Yukawa interactions with leptons
are introduced.
 It has been shown that
the simple models for masses of Majorana neutrinos
are excluded if $h \to \ell\ellp$ is discovered,
because constraints from $\ell \to \ellp\gamma$
cannot be evaded in such models.
 However,
we have also found that
there are five Mechanisms%
~(D3, D4, D11, D12 and D17 in Table~\ref{tab:D})
for masses of Dirac neutrinos
which can give a significant amount of $h \to \ell\ellp$
with the suppressed $\ell \to \ellp\gamma$ process.
 This is because
these models involve two kinds of scalar particles
($s_R^+$ and $s^{++}$, or $s_R^+$ and $s_2^+$)
which couple to $\ell_R$,
and then their contributions to $\ell \to \ellp\gamma$
can be cancelled with each other.
 In these Mechanisms,
Dirac neutrino masses are
generated as the following two forms,
$m_\text{D}^{} \propto y_\ell X_{SR}^\ast Y^s$
and $Y_\psi^\eta (Y_\psi^+)^\dagger Y^s$.
 Therefore,
future discovery of the nonzero $\BR(h \to \ell\ellp)$
shall be a strong probe of models for neutrino masses.
 Further probe is possible
if the second scalar (whatever it is neutral or charged)
is discovered in the current and future collider experiments
in addition to $h \to \ell\ellp$.


\begin{acknowledgments}
 The work of M.~A.\ was supported in part
by the Japan Society for the Promotion of Sciences~(JSPS)
Grant-in-Aid for Scientific Research~(Grant No.~25400250 and No.~16H00864).
 The work of S.~K.\ was supported in part
by Grant-in-Aid for Scientific Research on Innovative Areas,
The Ministry of Education, Culture, Sports,
Science and Technology,
No.~16H06492 and Grant H2020-MSCA-RISE-2014 no.~645722~(Non Minimal Higgs).  
\end{acknowledgments}

\end{document}